# A Peanut-hull-PLA based 3D printing filament with antimicrobial effect

Sabarinathan Palaniyappan [1], Narain Kumar Sivakumar [2], Ahmed S. Dalaq [1,3, *]

[1]Center for Biosystems and Machines, King Fahd University of Petroleum & Minerals, Dhahran, Saudi Arabia.

[2]Centre for Molecular Medicine and Diagnostics, Saveetha Dental College, Saveetha Institute of Medical and Technical Sciences, Saveetha University, India.

[3]Department of Bioengineering, King Fahd University of Petroleum & Minerals, Dhahran, Saudi Arabia.

Corresponding author *: ahmed.dalaq@kfupm.edu.sa

**Abstract:**

Peanut hulls, also known as *Arachis hypogaea L.* Particles (AHL), are an abundant biomass source with a long shelf life. In this study, we incorporate peanut hull powder into PLA polymer, imparting recyclability, biodegradability, and biocompatibility, along with the antimicrobial properties of AHL particles. In particular, we treat AHL particles as a reinforcement for PLA polymer to produce 3D printing filament compatible with the fused filament fabrication (FFF) 3D printing method. We provide a step-by-step method for preparing AHL particles, incorporating them into PLA, and ultimately forming high-quality filaments. We assess the quality of the filaments in terms of extruded dimensions, mechanical strength, and elastic modulus, along with physical properties such as porosity and melt flow index. We evaluate the printability and wettability of the filaments as well. Notably and unlike other biomass-based reinforcements in PLA, AHL preserves the filament's strength and enhances its elastic modulus. 3D-printed components fabricated using our PLA-AHL filaments successfully retain their antimicrobial properties and exhibit increased overall hardness. However, this comes at the expense of forming more microvoids, rougher surface, making the material more prone to fracture and leading to a slight reduction in fracture toughness with increasing AHL mass fraction.

## 1. Introduction

Complex topologies and geometrical features are now realizable through the combined use of computer-aided design and 3D printing technologies [1,2]. Both engineers and researchers alike are resorting to 3D printing for rapid prototyping, thereby bypassing expensive fabrication process typically used for intricate features, as well as time-consuming molding and casting techniques [3]. Among the various Additive Manufacturing (AM) technologies, Fused Filament Fabrication (FFF) has reached maturity, especially in the domain of polymer fabrication, and is the most widely used technology due to its key advantages, such as simplicity and cost efficiency [4].

Using FFF techniques, the quality of the end products (i.e., 3D prints) is strongly dependent on the type of raw materials used, whether granules or filament, the resolution of the 3D printer, and the structural features and overall geometry of the component to be printed. The component not only inherits the surface finish of the filament from which it is made but also its physical and mechanical properties, such as strength, toughness, elastic moduli, fracture toughness, hardness, overall density, and hydrophobicity/hydrophilicity [5,6]. The development of stronger, tougher filaments with a unique combination of physical, thermal, electrical, and potentially antimicrobial properties expands the range of applications and naturally endows printed components with desirable properties, enabling, for instance, smart actuation, damage resistance, modulated stiffness and their use in biomedical devices.

Industries have also focused on optimizing 3D printing filament through the lens of sustainability and the green economy. In particular, the adoption of biodegradable plastics is crucial, as continued use of non-degradable plastics results in high volume of solid waste [7,8]. Therefore, there is a pressing need to develop more sustainable 3D printing filament that is environmentally friendly and aligns with the aspirations of a complete circular economy [9]. Biopolymers appear to be a promising solution, as they incorporate naturally occurring organic particles that may maintain the structural and mechanical properties of the filament and in turn the 3D prints, while offering a unique combination of physical and perhaps biocidal properties [10,11].

Common polymer-based materials used in filament production include Polylactic Acid (PLA), Acrylonitrile Butadiene Styrene (ABS), Polyethylene Terephthalate (PET), Polyethylene Terephthalate Modified with Glycol (PETG), Polypropylene (PP), Polycarbonate (PC), and Polyamide (PA)[12]. PLA is a biodegradable, recyclable, and biocompatible thermoplastic biopolymer derived from corn starch [13]. However, PLA has a relatively low service life due to its limited strength and toughness [14,15].

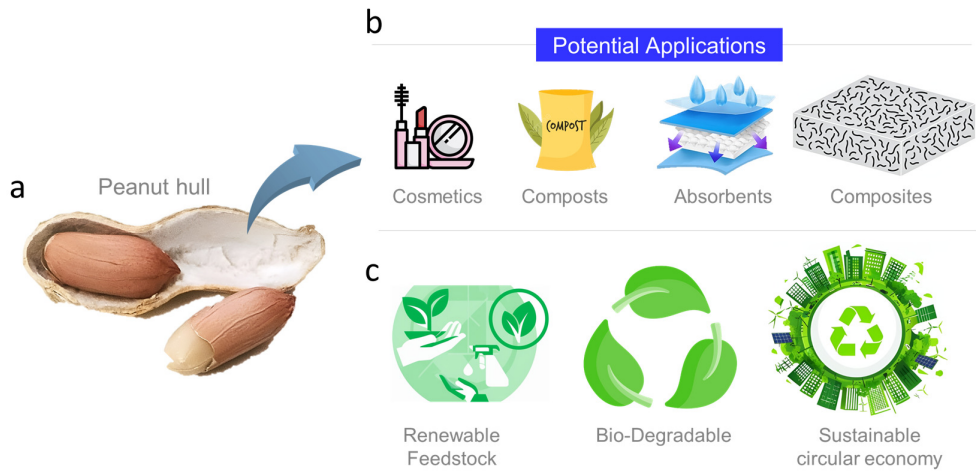

Figure 1. **A peanut shell and its potential application.** (a) typical peanut with shell, (b) potential applications of peanut shells in various industries, and (c) recycling of natural waste supports various sustainable practices.

As a mean to endow 3D printing filaments with enhanced physical properties or to improve their intrinsic mechanical properties, several investigators have resorted to natural biomass, such as flax [16], hemp [17], and coconut shell [18], as reinforcements for base polymers. The primary goal is to create lightweight products while ensuring recyclability and eco-friendliness of the final components. For example, the addition of walnut particles to a PLA matrix with the intention of making a more eco-friendly filament resulted in a decrease of both tensile and flexural strength by ~ 35% [19]. Similarly, the macadamia nut powder reinforcement in PLA reduces both tensile and flexural strength by 42% and 63%, respectively, particularly at high particulate mass fractions [20]. On the other hand, hemp fibres (a.k.a, hemp hurd) can enhance tensile properties, but only in moderation (<20% mass fraction), beyond which (>20%) the overall strength is compromised [21]. While biomass particles are widely available and abundant, incorporating them often comes at the cost of mechanical performance. Therefore, their use may not always be justified, considering the extrusion challenges and nozzle clogging issues encountered during filament production and 3D printing [22,23].

Among the many readily available, highly affordable, and abundant biomass sources, peanut hulls stand out. They have a long shelf life, lasting up to one year in their nut form and up to three years in powder form [24]. The peanut hull of interest here is obtained from *Arachis hypogaea L.* shown on Figure 1a, with an annual global production of a staggering 44 million tonnes. China is the largest producer, accounting for 45% of *Arachis hypogaea L.* (AHL) production, followed by India (16%) and the United States (5%) [25]. Chemically, AHL shells are rich in cellulose (37%), lignin (29%), and carbohydrates [26]. Despite their high carbohydrate content, they still exhibit considerable shelf life. It is the polyphenols, flavonoids, and isoflavones in AHL that give peanut hulls their antibacterial and antioxidant properties [27]. These affordable antimicrobial properties are what we seek to examine and incorporate into the 3D printing filament process. In fact, phenolic compounds are potent components in various clinical antimicrobial agents, spanning applications in cosmetics, absorbents, and composts (Figure 1b). Very few studies have reported the use of AHL particles in polymer composites [28,29]. To this end, this study explores the incorporation of AHL into PLA as a means to develop high-quality 3D printing filament based on AHL-PLA materials. In particular, we utilize abundant biomass in filament production, therewith enhance material properties, modulate hydrophilicity/hydrophobicity, and impart antimicrobial properties to 3D-prints. \We aim also to demonstrate the use of renewable feedstock like the peanut hulls in the production of 3D printing filament and ultimately in the fabrication of functional components. Hence, create an eco-friendly, recyclable, and fully biodegradable 3D printing filament based on PLA and AHL particles. In doing so, we demonstrate passing on AHL's potent polyphenols to 3D printing filament and in turn to 3D prints, moreover, assess the effect of adding AHL particles on the mechanical properties of the bio-composite. By incorporating such an abundant biomass source into the production of biodegradable raw filament for additive manufacturing space, we contribute to the broader sustainable circular economy initiative on a global scale (Figure 1c).

We begin our analysis with peanut hull (i.e., AHL powder) preparation and its incorporation into 3D printing filament for the FFF 3D printing technique. We conduct physical, dimensional, mechanical, and chemical characterization of the PLA-AHL-based filaments. Finally, we assess printability of PLA-AHL filaments, its hardness, toughness and most importantly whether the antimicrobial properties are successfully passed on to 3D-printed components and whether they, indeed, antimicrobial effect of AHL endured the multiphase changes involved in mixing, filament extrusion, and final 3D printing process.

## 2. A peanut shell-based 3D printing filament

*2.1 AHL-PLA based filament and 3D prints:*

Figure 2 shows the schematic procedure for extruding composite filaments from AHL particle synthesis to 3D printing of composite samples. Here we used Ingeo extrusion grade Polylactic acid (PLA) granules of 2003D as polymeric matrix. The polymer which has a density of 1.24 gm/cm$^3$, a melting temperature of 165–180°C and a relative viscosity of 4. The Leguminosae family of *Arachis hypogaea L* (Figure 2a) was selected as reinforcement for the study. Procured Arachis hypogaea L shells are cleaned with distilled water and dried in a vacuum at 70°C for 15 hours. After that the AHL particles are grounded in Thomas type Willys mill (RA Scientific Instruments, India), and then the fine powders are sieved using sieve shaker with sieve sizes of 240 (i.e., 240 opening per inch) mesh size. This sieved AHL particle (Figure 2b) are used as a reinforcement while the PLA polymeric granules as the matrix in the filament extrusion process (Figure 2c).

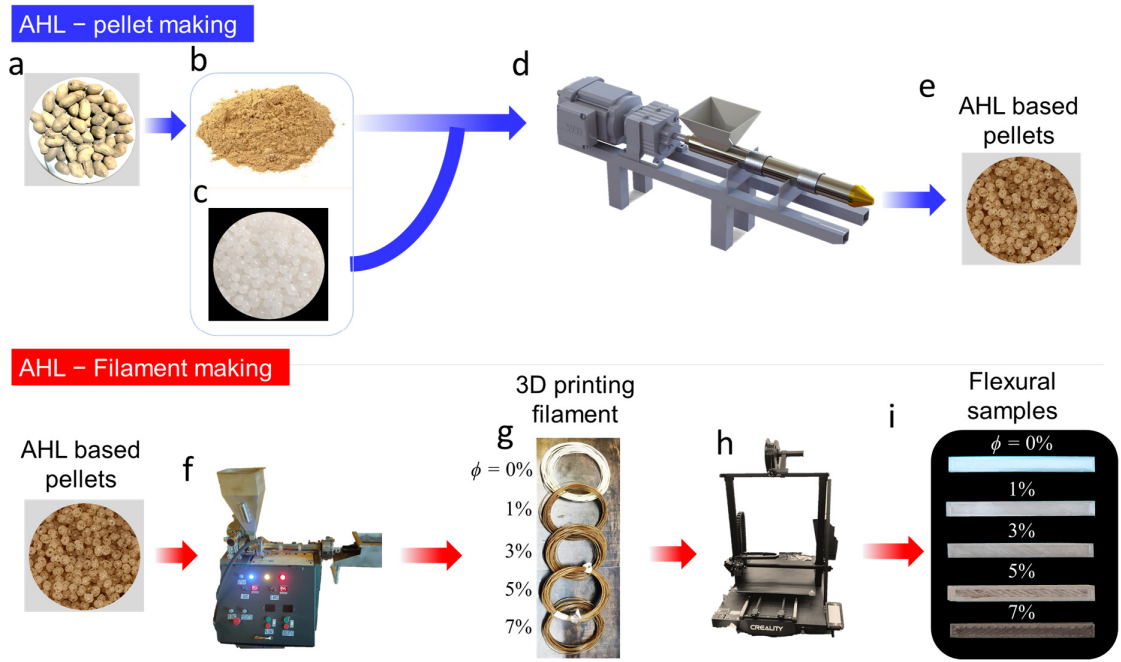

Figure 2. **Biocomposite filament production process.** (a) Raw peanuts (*Arachis hypogaea* L.) in (b) powder form. (c) Neat PLA pellets. All components are mixed in (d) an extruder, producing PLA-AHL-based pellets. The composite pellets are then transferred to (f) a filament extruder, which produces (g) 3D printing filament with varying levels of AHL particles, $\phi$. (h) The filament is used in an FFF 3D printer to produce standard test samples, such as (i) a flexural sample.

The filaments are extruded by the filament extrusion technique using filament extruder (Despar Integrators, India). Initially, the raw PLA pellets and a variable weight fraction of $\phi$ = 0, 1, 3, 5, 7 % of AHL particles are mixed together by dozing unit (Figure 2d). During dozing operation, the AHL particles and PLA polymeric granule mix are sprayed with an appropriate solvent that is chloroform, to improve the adhesion between the particles and granules. The blended AHL-PLA pellets (Figure 2e) are fed into the hopper and extruded with an extrusion speed of 40 mm/s, and extrusion temperature ranging from 155°C to 170°C. The extruded filaments are water quenched to obtain the uniform diameter of 1.75 ± 0.05 mm. Adding reinforcements beyond 7% AHL particle leads to various difficulties during extrusion such as clogging with blockage of particles in the extruder nozzle. Usage of two heating zone in the extrusion process limits the users up to $\phi$ = 7% of the composition for the extrusion.

The extruded AHL-PLA filaments with different weight percentage of AHL particle, $\phi$, is used as feedstock material for the 3D printing process. The different composition of AHL filaments (Figure 2g) were loaded in the fused filament fabrication (FFF) printer (Creality CR-10) to print the bio composite samples (Figure 2h). For the current study, the printer is attached with 0.8 mm nozzle, the printing temperature is 180 °C, and a bed temperature of 80 °C is maintained constant for all the samples. The extruded biocomposite filaments are 3D printed to produce Charpy, tensile, and flexural test samples (see Supplementary Materials), following their respective ASTM standards (see methods). The CAD models of the samples are sliced using CURA slicing software, with a printing speed of 20 mm/s, an infill density of 80%, a line infill pattern, and a layer height of 0.6 mm.

*2.2 Properties of AHL-PLA based filament*

The strength of the extruded composite AHL filaments is analysed using Universal testing machine (Instron, E3000) with a 5 kN load cell is attached with extensometer to measure the strain rate within the filament. Samples are prepared with a gauge length of 70 mm and the test was conducted as per the ASTM D4018 standard with a strain rate of 1 mm/min.

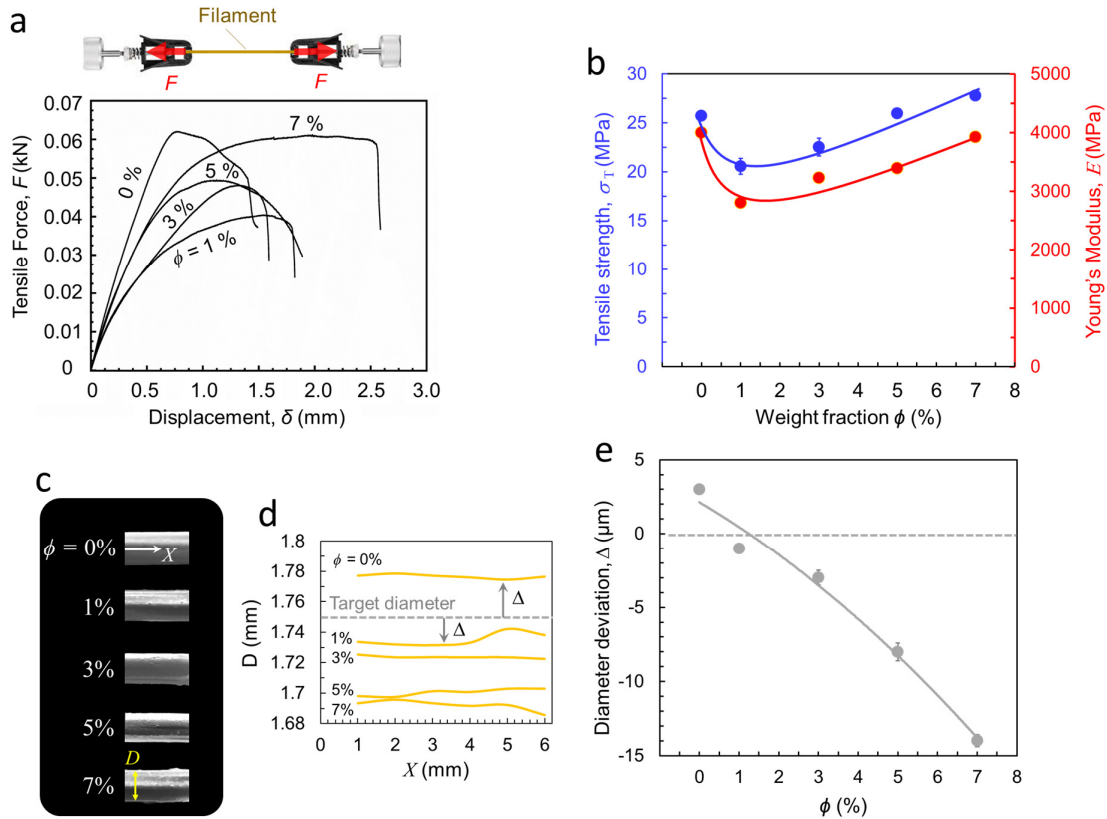

Figure 3. **Mechanical property results of the extruded AHL/PLA polymeric filaments.** (a) force-displacement plot, (b) Tensile results, (c) SEM images, (d) diameter deviation, and (e) mean diameter deviation of the extruded filaments.

Figure 3a illustrate the tensile test along with the respective representative force-displacement curves ($F - \delta$) for different AHL's weight ratios: $\phi$ = 0, 1, 3, 5, 7%. The $F - \delta$ curves show a more deformable behaviour with increased $\phi$. The maximum force, which reflects strength, drops with the introduction of the AHL particles (i.e., when $\phi > 0$) but increases gradually with further increases in $\phi$. At $\phi$ = 7%, the highest deformability with a maximum displacement of $\delta \approx 2.5$ mm is observed, indicating an improved ductility at high $\phi$. Figure 3b shows the corresponding mean tensile strength ($\sigma_T$, in blue) and Young's modulus ($E$, in red) of 3 samples for each $\phi$. The overall mean strength across different weight ratios is ≈23.8 MPa, with $\sigma_U$ fluctuating around this average. The maximum tensile strength of 27 MPa was observed at $\phi$ =

7%, indicating that particles may have been more effectively dispersed, leading to better load-sharing with the matrix. A similar finding was observed in crab shell-reinforced PLA filaments [30] and with the incorporation of *Trichosanthes cucumerina* fiber into the PLA matrix [31]. Similarly, $E$ drops at moment AHL particles are introduced ($\phi > 0$) and follows a linear increased in $E$ with increased $\phi$ reminiscent of simple linear correlation described by the rule of mixtures with increased $\phi$ [32]. Highest Youngs modulus of $E = 3.9$ GPa achieved at $\phi = 7\%$. AHL particulates appear to have a stiffening effect that also enhances strength with increasing $\phi$ when compared to low levels of AHL particles, $\phi = 1\%$. However, in relation to neat PLA, both Young's modulus and strength returned to the values of neat PLA ($\phi = 0$) when $\phi = 7\%$. An abrupt change in the extrusion process with the introduction of AHL particles for $\phi > 0$, can introduce imperfections between the AHL particles and the matrix, resulting in decreased Young's modulus and tensile strength.

Deviations in diameter of the extruded filament is analysed using Scanning Electron microscopic (SEM) technique (JSM –IT800 Nano SEM). An average diameter measurement of 30 readings were taken from each extruded filament. Variations in filament diameter was obtained at a 1 mm interval along its length. Figure 3c shows the SEM images of the extruded AHL−PLA biocomposite filaments. The diameter deviation of the filaments with various concentrations of $\phi = 0, 1, 3, 5,$ and 7% AHL particles is shown on Figures 3d and 3e. An increase in $\phi$ results in deviations, $\Delta$, from the target diameter (1.75 mm) of the filaments (i.e., diameter of extrusion nozzle), with the maximum diameter deviation observed in the filament having $\phi = 7\%$. All extruded filaments show nonuniformity on their surfaces with slight fluctuations in the diameter profile. As $\phi$ increases, the mean diameter, $D$, decreases and fluctuations gets pronounced (Figure 3d).

Figure 3e shows the deviation from the target diameter (1.75 mm): $\Delta = D - 1.75$ mm measured at 10 points along the filament. A reduction in diameter from the target size is observed which increases with the addition of AHL particles. These particles appear to impart irregularities to the filament surface, creating crests and valleys. Such surface irregularities are absent in the neat PLA filaments. The lowest diameter deviation was observed at $\phi = 1\%$, which is comparable to that of the neat PLA filament. A similar finding reported a decrease in the filament's diameter with the addition of fish scale powder [33]. Similar to other bioparticles, AHL particles are hydrophilic and possess hygroscopic properties [34–37], which can lead to bubble formation during the heated extrusion process. Formed bubbles may disrupt melt viscosity and, thus, affect the flow of the polymer composite resulting in variations in the extruded diameter.

## 3. Physical and chemical properties of 3D prints

Figure 4a shows the Fourier Transform Infrared (FTIR) spectra of the extruded neat PLA and AHL-PLA bio composite samples. The characteristics peak of PLA includes 2991 $cm^{-1}$ and 2935 $cm^{-1}$ indicates the C−H stretching of aliphatic groups [38]. The peak present at 1745 $cm^{-1}$, 1076 $cm^{-1}$, 1039 $cm^{-1}$, and 1181 $cm^{-1}$ is an indication of symmetric and asymmetric vibrational stretch of C−O−C bonds and bending of C−O group [39,40]. In case of AHL added PLA bio composite samples, the same characteristics peaks were present at 2991 $cm^{-1}$, 1745 $cm^{-1}$, 1076 $cm^{-1}$. For the AHL bio composite, the peak present at 1750 $cm^{-1}$ and 1591 $cm^{-1}$ is an indication of C = O stretching vibration of carbonyl group of hemicellulose and lignin [41]. The lignin present in the AHL particles which increases the covalent bond with cellulose and act as natural coupling agent between matrix and AHL fillers [42]. The peaks present in the region of 865 and 754 $cm^{-1}$ is an indication of amorphous and crystalline phases of 3D printed PLA [43].

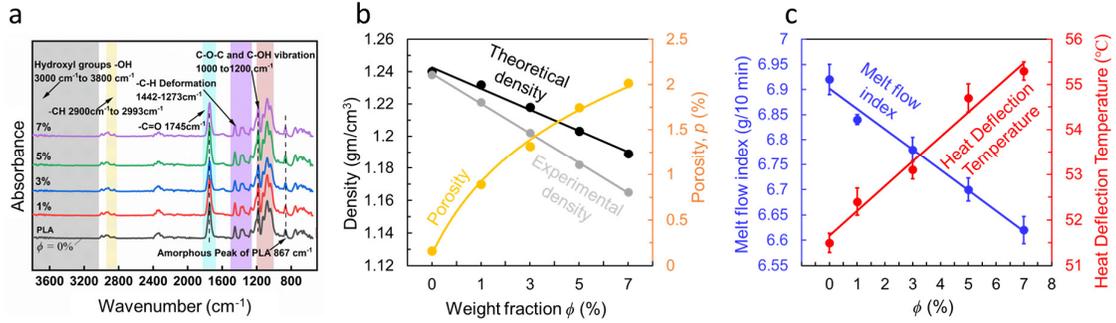

Figure 4. **Chemical properties and physical properties of the AHL-PLA filaments.** (a) FTIR spectra and (b) physical properties and (c) thermal properties of the neat PLA and AHL-PLA bio composite. The porosity in this plot is in percentage.

The addition of AHL particles does not produce any new peaks, indicating that the mixing and heated extrusion process did not induce any chemical or physical interactions that could form new functional groups or bonds. The decreases in the peak intensity at 1745 cm$^{-1}$ with respect to increases in the AHL percentage $\phi$. This is because of interaction of carbonyl group from the PLA polymer with the hydroxyl group of the added AHL particles, as seen in the use of organic particles in PLA [44,45].

To assess the printing accuracy in terms of 3D print's physical properties (Methods: Physical characterization). Figure 4b shows the theoretical and experimental density as well as the porosity of the 3D-printed neat PLA and AHL−PLA polymer composites. Both theoretical and experimental densities decline with increasing $\phi$, as AHL particles (having 0.77 g/cm³) are less dense than PLA (1.24 g/cm³). The difference between the theoretical and experimental densities is less than 2% indicating that despite surface irregularities and the hydrophilic nature of AHL, the particles were well-dispersed within the 3D-printed samples. Figure 4b shows that porosity percentage, $p\%$ increases with increased $\phi$. With higher $\phi$, the 3D print's surface becomes rougher due to the agglomeration of AHL particles which affects adhesion between the deposited filament layers. This, in turn, induces voids in the 3D-printed composite samples compared to neat PLA samples, a phenomenon also observed in PLA composites filled with wood particles [46].

Figure 4c shows the melt flow index (MFI) (in blue) and heat deflection temperature (HDT) (in red) results of the neat PLA and AHL-PLA polymer composites. Increasing $\phi$ decreases the MFI of the composite filaments. As the concentration of AHL particles increases, they act as obstructions to the melting PLA and disrupt the bonds within the PLA chains. Additionally, the free molecular movement of the polymeric chains in the AHL−PLA polymer composite becomes restricted. This behaviour is quite common and has been observed in several instances where organic particle reinforcements, such as wood particles in an HDPE matrix, decreased the MFI [47].

The heat deflection temperature (HDT) measures the thermal sensitivity of the developed composites which infers its service temperature limits. Figure 4c shows the HDT values with increased $\phi$, which indicates the softening point and thereby determines the upper and lower temperature limits for its working environment and so the suitable applications. HDT increased with $\phi$, implying that AHL particles can actually improve HDT of printed components. Highest HDT of 55.3°C occurs at $\phi = 7\%$. Therefore, incorporation of AHL particles enhances the thermal insulating capacity of the 3D-printed parts. In general, PLA softens under heat. The addition of AHL particles reduces this softening with increased temperature, thereby enhances thermal resistance and expands the potential applications of AHL-PLA composites compared to neat PLA. Other studies experimented with wood particle in PLA showed an increase in HDT as well [48].

## 4. Quality of the AHL-PLA 3D prints

Figure 5a shows the macroscopic images of 3D-printed AHL-PLA sample surfaces with different $\phi$. Initial visual and tactile examination of the 3D prints indicate that an increase in $\phi$ seems to induce more micropores. This is not observed in neat PLA with $\phi = 0$. In the case of neat PLA samples, micropores are indeed present at the interfaces between layers. However, when $\phi > 0$, with the addition of AHL particulates, the formation of micropores along the 3D printing lines within individual layers becomes more likely. This could potentially make the samples susceptible to damage as these micropores act as stress risers and thus initiation points for crack propagation along the filaments and printed lines/layers. These micropores also create space for moisture to diffuse into thereby increasing the wettability.

Figure 5b shows the 3D surface profile results of the 3D-printed AHL-PLA samples. Clearly, the increase in micropores with higher $\phi$ made the overall surface rougher with significant fluctuations, especially for $1\% < \phi < 5\%$. For $\phi = 7\%$, the surface became wavy with high amplitudes reaching up to ~576 nm and less frequency. Figure 5c illustrates the 3D average area surface roughness, $S_a$, of the AHL-PLA 3D prints (Methods). An increase in $\phi$ roughened the surface by ~10% for every 1% increase in $\phi$. In case of 5% and 7% there is a drastic increase in the surface roughness. The highest surface roughness was achieved at $\phi = 7\%$, measuring $S_a = 222.38$ nm. In addition to the increased prevalence of micropores with higher $\phi$, some unmelted PLA particles appear to have deposited on the surface, contributing to its waviness.

Figure 5d shows the SEM images of the fractured surfaces of neat PLA and AHL-PLA printed samples with increasing $\phi$. A triangular-shaped void is present between the layers in all cases, including neat PLA. The occurrences and mean size of these voids are lowest for the neat PLA sample and increase with $\phi$ (Figure 5d, e). The size of the microvoids is indicated by yellow arrows, as shown in Figure 5d. These interlayer voids are attributed to the turbulent flow of the polymer melt emerging from the nozzle, the infill process, and the contrasting cooling rates of the two materials involved (i.e., AHL and PLA), which formed air bubbles [38,49]. The failure mechanisms for the 3D-printed composites include matrix cracking (indicated a yellow shaded region for $\phi = 5\%$), layer-layer matrix interaction with friction, particle debonding, and particle pull-out from the matrix [50]. Additionally, in samples with 5% and 7% composition, minor pits are observed, indicating signs of particle pull-out.

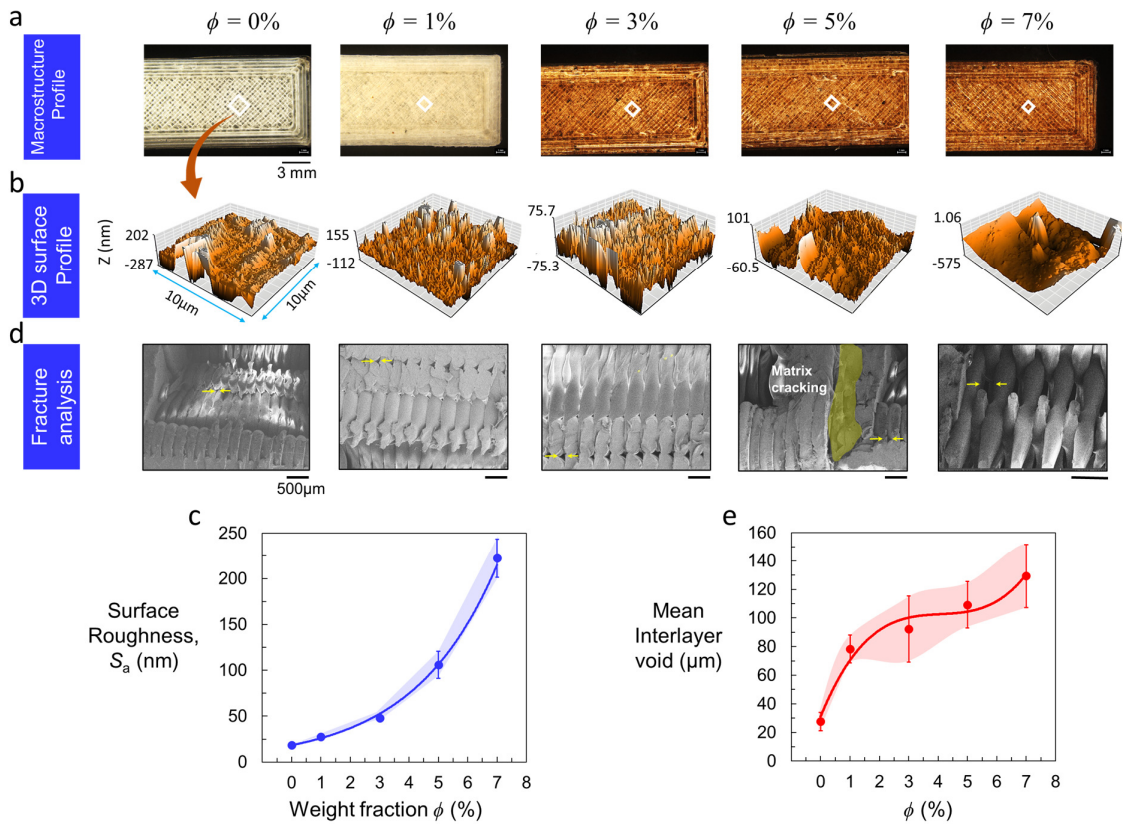

Figure 5. **Quality of 3D printed sample examination.** (a) Macrostructure, (b) 3D Surface roughness profile, (c) average area surface roughness value, (d) fractography of the tensile fractured sample and (e) interlayer voids of fractured tensile 3D printed AHL bio polymer composite. The shaded are indicates the variability (i.e., maximum and minimum errors) among samples.

## 5. Peanut-hull particles (AHL) makes PLA antimicrobial

It is known that unshelled peanuts have a shelf life of ~ 9 months without signs of bacterial infestation or fungal overgrowth, which encouraged us to incorporate them into the PLA matrix [51]. Generally, most bacteria prefer hydrophobic surfaces for adhesion, which promotes the formation of biofilms [52]. We have therefore used contact angle measurements to assess the wettability of the proposed AHL-PLA composite and assess its hydrophobicity/hydrophilicity. Figure 6a, b shows the contact angle, $\theta$, with increasing $\phi$, along with an image of a water droplet on the surface of the AHL-PLA composite shown on the right. The neat PLA already shows a hydrophilic response where $\theta < 90°$. The angle $\theta$ decreased with increased $\phi$ down to 61.66°. Specifically, the contact angle values are 66.45°, 65.58°, 64.48°, and 61.66° for $\phi$ = 1%, 3%, 5%, and 7%, respectively. Clearly, the introduction of AHL particles ($\phi > 0$) makes the composite more hydrophilic. This is because AHL particles, like other plant-based shells (e.g., walnut), are naturally hygroscopic and have a high affinity for water molecules [34–37,53]. When the 3D-printed AHL biocomposite samples come into contact with water, the AHL particles show tendency of spreading the water droplet over the surface. We measured the water absorption by immersing the sample in normal water and the water absorption tendency of the samples are measured after 24 hours (see Methods). Figure 6c shows the water absorption percentage with increased $\phi$, which shows a linear increase of water absorption with $\phi$. The PLA-AHL composite absorbs 0.12 % of water for every 1% increase $\phi$. For comparison, introduction of 0 – 10% mass fraction of Agav fibres in PLA results in only 2 – 4% increase in water absorption with respect to neat sample [54]. Moreover, adding 3% untreated *Arundo donax* L. as a reinforcement to PLA yielded a water absorption of ~1.3%, which makes it sensitive to moisture and comparable to PLA-AHL composites [55]. One possible reason for the increased hydrophilicity with higher AHL particle content is the natural presence of functional groups such as hydroxyl (-OH), carbonyl (C=O), and carboxyl (-COOH) in the hemicellulose and lignin [56]. These groups facilitate the formation of hydrogen bonds with water molecules that makes the particles more hydrophilic. Additionally, the incorporation of AHL particles induces more micropores as shown in Figure 5, which provide diffusion sinks that draw water toward them. Fabricated 3D prints with higher water absorption characteristics are prone to swelling of its layers and thus become prone to layer peel off before its service life period. Similar finding was reported by wood, bamboo, and cork-based wood reinforced PLA composites shows increases in the water uptake by increasing the concentration of such organic wood particles in the PLA matrix [57]. Overall, the proposed AHL-PLA prints exhibit hydrophilicity, making them moisture-sensitive which can weaken polymer chains, reduce strength, and may cause premature degradation in structures and mechanical components. However, those attributes are beneficial for biodegradability, biomedical filtration, tissue engineering and drug delivery and compostable applications.

We tested whether 3D-printed samples made from AHL-PLA filament possess any degree of antimicrobial effect. A 3D printed small disk sample having a 6 mm in diameter (20 µg/mL) was subjected and tested against Staphylococcus aureus, a known Gram-positive bacterial strain that is potentially pathogenic and commonly infests biomedical devices [58] (Figure 6d). Figure 6e shows the inhibition zone diameter (indicated by the circle enclosed by two red arrows) for different levels of $\phi$. The highest inhibition zone of 14.3 mm was measured at $\phi = 7\%$. Clearly, increased AHL particle content inhibits bacterial growth on and around the composite sample, effectively preventing the growth of Gram-positive bacteria like Staphylococcus aureus. The phenolic compounds, lectins, and flavonoids present in the AHL particles can damage bacterial cell membranes and hinder bacterial attachment to the sample surface. More specifically, those compounds can cause a loss of membrane permeability [59]. To put the antimicrobial performance of the AHL-PLA composite into perspective, we compare its inhibition zone with that of onion (150 mg/mL) and garlic peels (150 mg/mL) [60], cashew nutshell (78.1 μg/mL) [61], and pecan nutshell (250 to 31.25 mg/mL) [62], all against *Staphylococcus aureus*. These are raw shells with known antimicrobial effects, with varied densities/concentrations (ml/ml) that often exceed that of our AHL-PLA (i.e., >20 μg/mL) 3D-printed sample. Yet, our AHL-PLA 3D-printed sample showed comparable, if not superior antimicrobial resistance even in its final 3D-printed form. Nevertheless, AHL-PLA has not yet approached the effectiveness of potent antibiotics such as ciprofloxacin and gentamicin (with disc potency of 10μg [63]) which respectively possessed 74% and 47% larger inhibition zones than that of our AHL-PLA sample (at $\phi$ = 7%). The compared inhibition zones are based on Clinical and Laboratory Standard Institute (CLSI) standards, where all tested are performed on 6 mm disk sample.

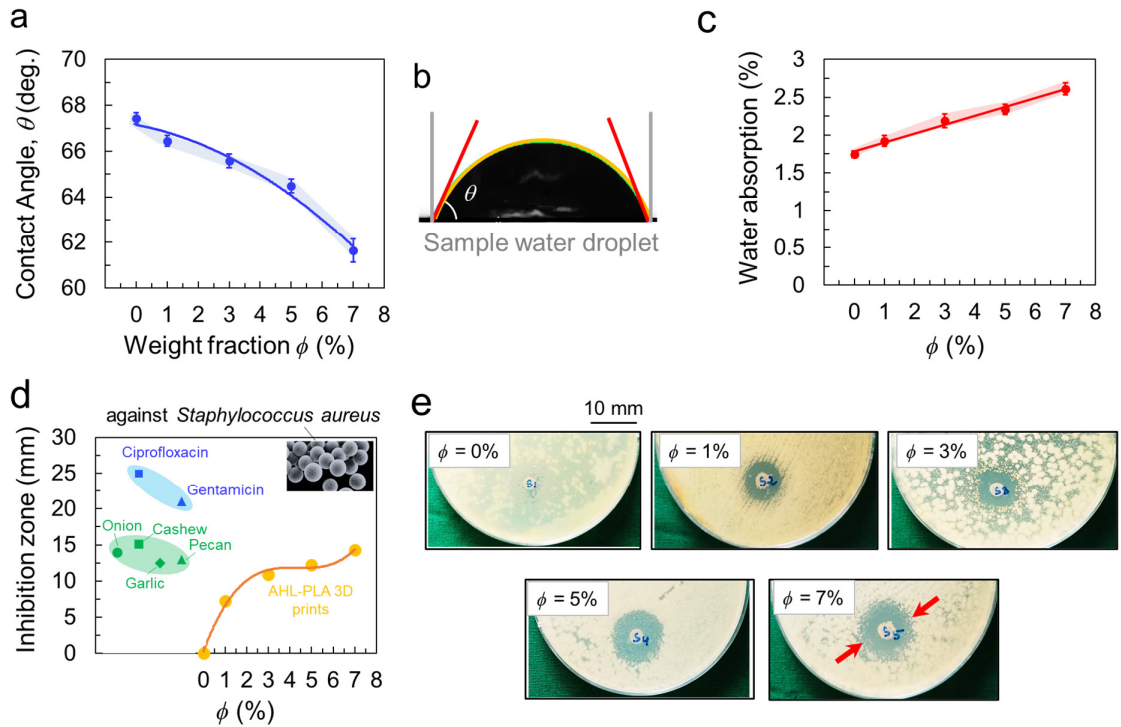

Figure 6. (a) Contact angle and its corresponding (b) droplet image, (c) water absorption, and (d) inhibition zone as a function of $\phi$, with bacterial strain images of 3D-printed AHL-PLA. (e) Inhibition zone measurement (enclosed by the red arrows) of a 6 mm 3D-printed disk.

## 6. Toughness of AHL-PLA samples

We have already reported the tensile strength and Young's modulus of AHL-PLA filaments with increasing AHL content, $\phi$, in Figure 3. Here, we examine how variations in Young's modulus and tensile strength affect 3D-printed samples. Preliminary tests indicate that tensile strength, flexural strength, and Young's modulus decrease with increasing $\phi$ in the 3D-printed samples (Supplementary Material). This contrasts with the increasing trends observed for tensile strength and Young's modulus in the filament samples earlier on Figure 3. As previously mentioned, the 3D-printed samples contain micropores which leads to an overall decline in strength and Young's modulus by compromising the microstructure. Figure 7a shows the Shore D hardness levels as a function of increasing $\phi$, indicating that hardness increases with $\phi$. Hardness is generally known to increase with Young's modulus $E$, but not necessarily with its bulk values, which are often influenced by microstructural features and microvoid volume fractions, thereby affecting localized hardness measurements.

We present here the impact fracture toughness (dynamic critical energy release rate, $G_c$) of the sample rather than its Young's modulus. In other words, we focus on its resistance to impact loading or sudden falls, as might occur in actual biomedical devices or engineering components. Figure 7b shows the Charpy impact tester (ASTM D6110), where three V-notched samples were 3D printed using AHL-PLA filaments at each $\phi$. Figure 7c shows the impact fracture toughness of the samples. Due to imperfect bonding between layers and the increased microvoids associated with higher $\phi$, the $G_c$ of the 3D-printed samples decreases following a linear decline with increasing $\phi$ which is shown in Figure 7d. The geometric nature of microvoids can be inferred from simple micromechanical damage models [64]. From a purely geometric micromechanical perspective of porous microstructures, the relative toughness takes a simple form with porosity $p$:

$$G_c / G_c^{PLA} = 1 - p^n = 1 - p^{0.5} \tag{1}$$

The exponent $n \approx 0.5$ was found using least square mean error curve fitting technique [65]. $G_c^{PLA}$ denotes the fracture toughness of the neat PLA, that is $G_c^{PLA}$ = 14.4 kJ/m². $p$ here is the porosity of the filaments from which the 3D printed sample is made which range from: $p = [0, 1]$. Indeed, the porosity of the actual 3D printed sample can be larger than the $p$. The exponent $n$ can roughly reflect the geometric features of microstructures. In this case, since $n$ is close to 2/3, it resembles a closed porous microstructure with some broken ligaments [64].

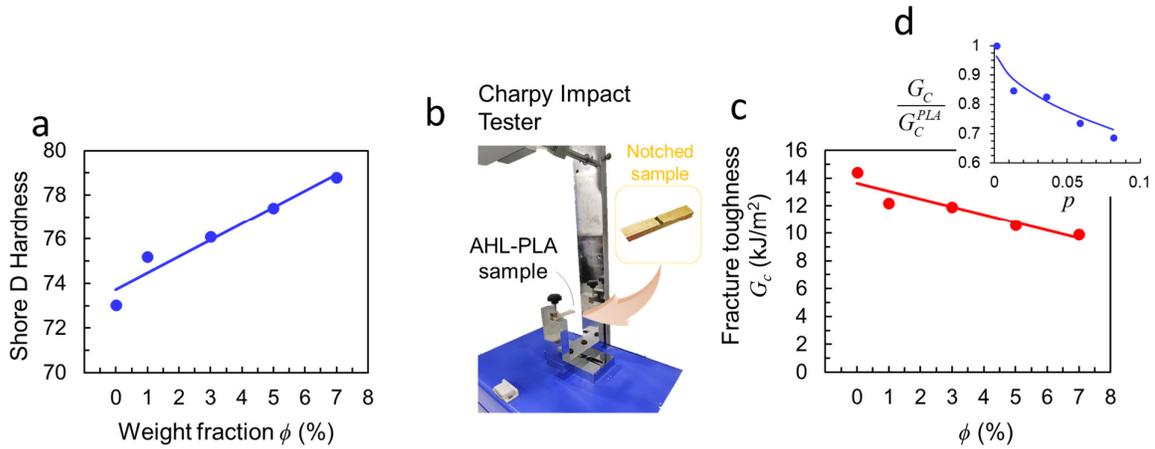

Figure 7 Mechanical properties of the 3D printed AHL/PLA composites (a) Shore D hardness, (b) Charpy impact tester and (c) dynamic fracture toughness with AHL concentration $\phi$, (d) shows the relative fracture toughness $G_c / G_c^{PLA}$ as function of porosity $p$ (range from $p = [0, 1]$).

The added AHL reinforcements, which are rich in hemicellulose and pectin (hard particles contributing to increased hardness) clearly embrittle the PLA matrix [66]. Other factors, such as particle dispersion, and the wetting of the added particles, could also be responsible for the reduction in the impact strength of the 3D-printed composites [67]. The degree of crystallinity in the polymer affects its mechanical performance; higher crystallinity can increase stiffness but may also make the material more brittle thereby decreasing impact strength. There is a 30% drop in dynamic fracture toughness at $\phi$ = 7%, from 14.4 kJ/m² to 9.9 kJ/m² compared to neat PLA. Indeed, caution is required when introducing AHL particles, as they seem to embrittle the 3D-printed components.

**Conclusion**

This study focuses on the development of biopolymeric composite filament made of peanut shell particles (*Arachis hypogaea* L., AHL) as reinforcement in poly(lactic acid) (PLA) polymer. We have outlined a step-by-step procedure for making such peanut hull-based filaments. We demonstrate its 3D printability and assess filament mechanical properties, extrusion accuracy, chemical composition, overall density, porosity, melt flow index, and heat deflection temperature with varying AHL particle weight fraction $\phi$. We carried out scanning electron microscopy (SEM) imaging of 3D-printed samples along with surface profilometry measurements. Finally, we examined whether the antimicrobial properties of AHL sustain the extrusion and 3D printing process and whether it maintained its effectiveness in the final 3D-printed samples.

Peanut shells can be extracted, ground, mixed with PLA pellets, and extruded into AHL-PLA pellets with varying $\phi$. The AHL-PLA filament in turn is used in a filament extruder to produce AHL-PLA filament with different $\phi$ values. The peanut-based filaments can finally be used to create AHL-PLA based 3D prints.

We found that introducing AHL into PLA makes the filament more deformable. While the filament's tensile strength and Young's modulus initially decrease, they quickly become stronger and stiffer with increased AHL particle content. The filament's diameter deviates from the target diameter of the extrusion nozzle and tends to decrease (becomes thinner) with increased $\phi$. The filament's density declines with $\phi$, while porosity increased with increased $\phi$.

The AHL-PLA composite retains the same chemical composition as AHL and PLA, with no additional compounds or functional groups emerging from the extrusion process.

With higher AHL particle content, the melting and flowability of the AHL-PLA composite decrease while its thermal sensitivity also declines (as indicated by a decrease in heat deflection temperature, HDT, with increasing $\phi$). As such, 3D printability decreases with increased AHL particles and thus limits were found to be at $\phi = 7\%$. Due to increased porosity resulting from the introduction of AHL particles, the surfaces of 3D-printed samples made from AHL-PLA filament tend to become rougher at higher $\phi$, with more and larger microvoids and interlayer gaps. This makes the 3D prints more susceptible to fracture due to inherent structural damage.

Interestingly, the antimicrobial properties attributed to the polyphenols, flavonoids, and isoflavones present in the AHL particles were preserved and sustained throughout the multi-phase and heated dual-extrusion process. The 3D prints exhibited a considerable inhibition zone that is comparable to raw known antimicrobial shells such as garlic and onion and up to 60% of that of gentamicin against *Staphylococcus aureus*.

All in all, the 3D-printed samples become harder due to AHL particles, but the increased microvoids and porosity add damage to filament and in turn to 3D prints. This study demonstrates that biodegradable 3D-printed samples with reliable mechanical properties can be produced using AHL-PLA filaments, yielding microbial-resistant, harder, rougher surfaces with slightly brittle behaviour. The appearance of the 3D-printed AHL-PLA composite resembles wood which makes it suitable for wood replacement applications. Possible applications include but not limited to furnisher, toys, door handles, kitchen utensils, boxes, biomedical instruments and food packaging applications. As food packaging requires antimicrobial property to avoid bacterial and fungal contamination.

**Methods**

**ATR-FTIR Analysis**

Fourier Transform InfraRed (FT-IR) spectroscopic technique was used to analyse the functional modification of the PLA and the respective AHL/PLA bio polymer composite. The test was conducted in the FTIR equipment (Bruker alpha FTIR spectrometer) in ATR mode with the frequency ranges from 400 – 4000 cm$^{-1}$ with spectral resolution of 4 cm$^{-1}$.

**Physical characterization**

The theoretical density of the 3D printed neat PLA and AHL/PLA bio composite sample is measured using the rule of mixture technique and which is calculate using the following equation (1S)

$$\rho_{th} = \frac{1}{\frac{W_P}{\rho_P} + \frac{W_R}{\rho_R}} \qquad (1S)$$

Whereas W$_P$ and W$_R$ represents the weight fraction of AHL particle and PLA polymer and $\rho_P$ and $\rho_R$ represents the density of AHL particle and PLA polymer, respectively.

The experimental density of the 3D printed neat PLA and AHL-PLA bio composite sample is measured using Archimedes principle. Based on the initial and final mass of the samples in air and water, and density of water is used to determine the experimental density of the 3D printed AHL-PLA bio composite. The experimental density is calculated based on the equation (2S) mentioned below,

$$\rho_{exp} = \frac{W_a \rho_w}{W_a - W_w} \qquad (2S)$$

Where the $W_a$ and $W_w$ denote the weight of the sample in air and water and $\rho_w$ is the density of water which is 1g/cm$^3$.

The porosity of the prepared 3D printed neat PLA and AHL/PLA bio composite is measured using ASTM D2734 by normalising the theoretical and experimental density of the composite.

Melt flow index of the extruded filament is used to measure flowability of the composite melt for the 3D printing operation. The experiment was done in a Meltfixer LT equipment, with an extrusion load of 2.16kg and temperature of 180°C. Each condition 10-time samples were extruded for all the compositions and the average value is considered as melt flow index value.

Macroscopic observation of the 3D printed samples is observed using Stereo microscope (Leica M205 C, US) and the observation was made on the resolution of 0.952 µm. Surface roughness of the AHL-PLA printed samples is measured using Atomic Focal Microscopic technique (Ultra nanoindentation tester, UNHT, Anton Paar, Germany) to show the 3D surface profile. The measurement was made over a square area of 10 x 10 μm$^2$ in a static mode condition and the surface profile was analysed using the NanoSurf C3000 software [68].

Areal average surface roughness, $S_a$, parameter is measured to analyse the rough and smooth profiles along the selected profile. The Areal average surface roughness $S_a$ is calculated based:

$$S_a = \frac{1}{A} \iint_A |z(x,y)|\, dxdy \tag{3S}$$

Whereas $A$ indicates total surface area measured, and $z(x,y)$ indicates height of the 3D profile in the mean plane.

**Mechanical analysis**

Mechanical properties such as hardness and impact test of the neat PLA and AHL-PLA reinforced bio composite samples are done as per the ASTM standards. Shore hardness of the AHL bio composite samples is measured to understand the resistance under deformation of the material by the addition of AHL particle into the polymer matrix. The experiment was done in the shore hardness tester (CASA, India), as per the ASTM D2240 standard. For each sample 12 measurement were taken at 12 different places and its average value is taken as the shore hardness of the respective configuration. Impact fracture toughness (dynamic energy release rate) of the samples is measured to estimate the energy absorption of the developed AHL-PLA 3D prints. The test was done on the Charpy impact test machine based on the ASTM D6110 standard. The samples are designed and 3D printed with cross section of 126 x 12.7 x 3.4mm. During the impact test, the hammer was strike at a speed of 2.9 m/s and the impact strength is calculated based on the difference between the absorbed energy with respect to the width and height of the samples.

**Water absorption and contact angle test.**

Swelling behaviour of the organic AHL particle reinforced polymeric samples are measured using water absorption test. For the current study, the 3D printed AHL/PLA bio composite samples with different combinations are measured using the ASTM D570-99 standard. 3D printed samples with dimension of 39 x 10 x 3.4mm is used for the study. Samples were dried and the initial weight is measured and the samples were dipped in water for 24 hours, further its wiped with dry tissue paper and weighed the final weight of the samples. Water absorption of the AHL/PLA bio composite samples is measured based on the variation of mass difference of the samples and it's calculated by the equation (4S) mentioned below:

$$\text{Water absorption} = \frac{W_f - W_i}{W_i} \times 100 \qquad (4S)$$

Whereas $W_i$ and $W_f$ indicate the initial and final weight before and after immersion of 3D printed AHL-PLA composite samples in normal water. Contact angle measures the surface phenomenon of the 3D printed bio composites under capillarity mode of water absorption. The experiment was conducted in the Ossila contact angle goniometer under sessile drop technique [69]. Samples with different composition of AHL particles is 3D printed and the distilled water of 5µL is placed horizontally on the 3D printed surface by syringe. For each sample three images were taken and the angle is measured in both right and left side and its average value is taken for the contact angle values of the respective composition.

**Heat deflection temperature**

Heat deflection temperature of the 3D printed AHL-PLA composite sample is tested in the Presto HDT/VSP tester under the three point bending conditions. The test was conducted under ASTM D648 standard with cross section of 125mm × 12.5mm × 3mm. The measurement for the heat deflection was measured at a load of 0.5 MPa under heating rate of 2°C.

**Anti-microbial study**

Staphylococcus aureus as Gram-positive bacteria and it is selected to analyse the anti-bacterial activity of the different composition of AHL/PLA bio composite by Agar diffusion method. Initially the bacteria were cultured in the lysogeny broth (LB) and sterilised, incubated for 4 hours at 37°C. 50 µL of the bacterial suspension was evenly spread on the LB Petri dish plate. The bio composite samples are placed on the petri dish plate and incubated overnight at 37°C. Based on the obtained macroscopic images, the inhibition zones are measured using ImageJ software.

**Morphological analysis of 3D printed AHL bio composite samples.**

Surface phenomenon, microstructural characterization and AHL particle distribution in the PLA matrix was analysed using Scanning Electron Microscope (Make: JSM –IT800 Nano SEM). Further mode of fracture and damage mechanisms were also evaluated on the tensile tested samples.